\documentstyle[osa,manuscript,graphicx]{revtex}

\begin{document}

\def\lsim{\ ^<\llap{$_\sim$}\ }
\def\gsim{\ ^>\llap{$_\sim$}\ }
\def\r2{\sqrt 2}
\def\beq{\begin{equation}}
\def\eeq{\end{equation}}
\def\beqn{\begin{eqnarray}}
\def\eeqn{\end{eqnarray}}
\def\rmuu{\gamma^{\mu}}
\def\rmud{\gamma_{\mu}}
\def\PL{{1-\gamma_5\over 2}}
\def\PR{{1+\gamma_5\over 2}}
\def\sinW2{\sin^2\theta_W}
\def\AEM{\alpha_{EM}}
\def\mul{M_{\tilde{u} L}^2}
\def\mur{M_{\tilde{u} R}^2}
\def\mdl{M_{\tilde{d} L}^2}
\def\mdr{M_{\tilde{d} R}^2}
\def\mz2{M_{z}^2}
\def\c2b{\cos 2\beta}
\def\au{A_u}
\def\ad{A_d}
\def\cob{\cot \beta}
\def\v#1{v_#1}
\def\tb{\tan\beta}
\def\epem{$e^+e^-$}
\def\KK{$K^0$-$\bar{K^0}$}
\def\wi{\omega_i}
\def\xj{\chi_j}
\def\Wmu{W_\mu}
\def\Wnu{W_\nu}
\def\m#1{{\tilde m}_#1}
\def\mH{m_H}
\def\mw#1{{\tilde m}_{\omega #1}}
\def\mx#1{{\tilde m}_{\chi^{0}_#1}}
\def\mc#1{{\tilde m}_{\chi^{+}_#1}}
\def\mwi{{\tilde m}_{\omega i}}
\def\mxi{{\tilde m}_{\chi^{0}_i}}
\def\mci{{\tilde m}_{\chi^{+}_i}}
\def\mz{M_z}
\def\sw{\sin\theta_W}
\def\cw{\cos\theta_W}
\def\cb{\cos\beta}
\def\sb{\sin\beta}
\def\rwi{r_{\omega i}}
\def\rxj{r_{\chi j}}
\def\rfp{r_f'}
\def\Kik{K_{ik}}
\def\Fq2{F_{2}(q^2)}
%

\begin{center}{\Large \bf  
 Effects of CP violation on Event Rates   
  in the Direct Detection of Dark Matter
 } \\
\vskip.25in
{Utpal Chattopadhyay$^a$, Tarek Ibrahim$^b$, \& Pran Nath$^c$  }

{\it
a Department of Physics, Indiana University, Bloomington, 
IN 47405, USA\\
b Department of Physics, Faculty of Science, University of 
Alexandria, Alexandria, Egypt\\
c Department of Physics, Northeastern University, Boston, MA 02115, USA\\
}

\end{center}

\begin{abstract}                
A full analytic analysis of the effects of CP violating phases 
on the event rates in the direct detection of dark matter 
 in the scattering of 
neutralinos from nuclear targets is given. The analysis includes
CP violating phases in softly broken supersymmetry in the
framework of the minimal supersymmetric standard model (MSSM) 
when generational mixings are ignored. 
A numerical analysis shows that large CP violating phases 
including the constraints from the experimental limits on the 
neutron and the electron electric dipole moment (EDM) can 
produce substantial effects on the event rates in dark matter
detectors.
\end{abstract}

 Supersymmetric theories with R parity conservation imply  the 
existence of a lowest mass supersymmetric particle (LSP) which is 
absolutely stable\cite{jungman}. In supergravity models\cite{applied}
 with R parity invariance one finds that over
a majority of the parameter space the LSP is the lightest neutralino,
and thus the neutralino becomes 
a candidate for cold dark matter (CDM) in the universe. A great deal
of dark matter exists in the halo of our galaxy and estimates of the
density of dark matter in our solar neighborhood give densities of
0.3GeV/cm$^3$ and particle velocities of $\approx 320 kms^{-1}$.
One of the interesting suggestions regarding the detection of dark
matter is that of direct detection  via scattering of neutralinos
from target nuclei in terrestrial detectors\cite{good}. 
Considerable progress has
been made in both technology of detection\cite{santa} as well as in accurate
theoretical predictions of the expected event rates in detectors such
as Ge, NaI, CaF$_2$, and Xe\cite{flores,drees,events,arno}.  

In this paper we discuss the effects of  CP violating phases in 
supersymmetric theories on event rates in the scattering of neutralinos
off nuclei in terrestrial detectors. Such effects are negligible if
the CP violating phases are small. Indeed the stringent experimental 
constraints on the EDM of the neutron\cite{altra} and of the 
electron \cite{commins} would seem to require
either small CP violating phases\cite{bern} 
 or a 
heavy supersymmetric particle 
spectrum\cite{na}, in the range of several TeV, to satisfy the experimental 
limits on the EDMs. However, a heavy sparticle spectrum also
constitutes fine tuning\cite{ccn} and further  a heavy 
spectrum in the range of several TeV will put the supersymmetric 
particles beyond the reach of even the LHC. 
 Recently a new possibility was proposed\cite{in1,in2,correction}, 
i.e., that of internal cancellations among the various components 
contributing to 
the EDMs which allows for the existence of large CP violating phases,
with a SUSY spectrum which is not excessively heavy and is thus accessible
at colliders in the near future. 
CP violating phases O(1) are attractive 
because they circumvent the naturalness problem associated with small
phases or a heavy SUSY spectrum. 
The EDM analysis of Ref.\cite{in1} was for the minimal supergravity model
which has only two CP violating phases. The analysis of 
Ref.\cite{in1} was extended  in Ref.\cite{in2} to take account of
all allowed 
CP violating phases in the Minimal Supersymmetric Standard Model (MSSM)
with no generational mixing. This extension gives 
the possibility of the cancellation mechanism to occur over a 
much larger region of the parameter space allowing for 
large CP violating phases over this region. Indeed a general numerical
analysis bears this out\cite{bgk}. 

 Large CP violating phases can affect significantly
   dark matter analyses and other phenomena at low 
energy\cite{falk1,bk,falk2,falk3}. A detailed analysis in Ref.\cite{falk2}
shows that large CP violating phases consistent with the cosmological
relic density constraints and the EDM constraints using the 
cancellation mechanism are indeed
possible. CP violating phases affect event rates in dark 
matter detectors and a partial analysis of these  
effects with two CP phases  and without the EDM
constraint was given 
in Ref.\cite{falk3}.  
Thus, currently there are no analyses where the effect of large
CP violating phases on event rates and the simultaneous satisfaction
of the EDM constraints via the cancellation mechansim are discussed.
Further, the previous analyses are all  limited to two CP violating 
phases while supergravity models with non-universalities and MSSM
can possess many more phases\cite{in2}.
In this paper we give the first complete analytic analysis of the effects of
CP violation on event rates with the inclusion of all CP 
violating phases allowed in the minimal supersymmetric standard 
model (MSSM)  when intergenerational 
mixings are ignored. We then give a numerical analysis of  the 
CP violating effects on event rates  with the inclusion of the EDM constraints.
It is shown that while the effects of CP violating phases on event rates
are significant with the inclusion of the EDM constraints, they are not
enormous as for the case when  the EDM constraints are ignored. 

  We discuss now the details of the analysis in MSSM. The nature of the
  LSP at the electro-weak scale is determined by the neutralino
  mass matrix which in the basis ($\tilde B$, $\tilde W$, $\tilde H_1$,
  $\tilde H_2$), where $\tilde B$ and $\tilde W$ are the U(1) and 
  the neutral SU(2) gauginos, is given by

\beq
\left(\matrix{|\m1|e^{i\xi_1}
 & 0 & -\mz\sw\cb e^{-i\chi_1} & \mz\sw\sb e^{-i\chi_2} \cr
  0  & |\m2| e^{i\xi_2} & \mz\cw\cb e^{-i \chi_1}& -\mz\cw\sb e^{-i\chi_2} \cr
-\mz\sw\cb e^{-i \chi_1} & \mz\cw\cb e^{-i\chi_2} & 0 &
 -|\mu| e^{i\theta_{\mu}}\cr
\mz\sw\sb e^{-i \chi_1} & -\mz\cw\sb e^{-i \chi_2} 
& -|\mu| e^{i\theta_{\mu}} & 0}
			\right)
\eeq
Here $M_Z$ is the Z boson mass, $\theta_W$ is the weak angle, 
$\tan\beta$=$|v_2/v_1|$ where $v_i=<H_i>$ = $|v_i|$ 
$e^{i\chi_i}$  (i=1,2) where $H_2$ is the Higgs that
gives mass to 
the up quarks and $H_1$ is the Higgs that gives mass to 
the down quarks and the leptons, $\mu$ is the Higgs 
mixing parameter (i.e. it appears in the superpotential as the 
term $\mu H_1H_2$), 
$\m1$ and $\m2$ are the masses of the U(1) and SU(2) gauginos at
the electro-weak scale with $\xi_1$ and $\xi_2$ being their phases. 
 The 
neutralino mass matrix of Eq.(1) contains several phases. 
However, it can be shown \cite{in2} 
that the eigenvalues 
and the eigenvectors of the neutralino mass matrix 
depend on only two combinations: $\theta$=
$\frac{\xi_1+\xi_2}{2}$+$\chi_1$ +$\chi_2$+$\theta_{\mu}$ 
and $\Delta \xi$=$(\xi_1-\xi_2)$.
The neutralino matrix can be   
 diagonalized  by a unitary matrix X  such that
\beq
X^T M_{\chi^0} X={\rm diag}(\mx1, \mx2, \mx3, \mx4)
\eeq
 We shall denote the LSP by the index 0 so that
 
 \beq
 \chi^0=X_{10}^*\tilde B+X_{20}^*\tilde W+ X_{30}^*\tilde H_1+
 X_{40}^*\tilde H_2
 \eeq
 The basic interactions that enter in the scattering of the LSP from 
 nuclei are the neutralino-squark-quark interactions in the s channel
 and the neutralino-neutralino-Z(Higgs) interactions in the cross channel.
 The squark mass matrix $M_{\tilde{q}}^2$
 involves both the phases of $\mu$ and 
 of the trilinear couplings as given below 
 
\beq
\left(\matrix{M_{\tilde{Q}}^2+m{_q}^2+M_{z}^2(\frac{1}{2}-Q_q
\sin^2\theta_W)\cos2\beta & m_q(A_{q}^{*}m_0-\mu R_q) \cr
   	          	m_q(A_{q} m_0-\mu^{*} R_q) & M_{\tilde{U}
}^2+m{_q}^2+M_{z}^2 Q_q \sin^2\theta_W \cos2\beta}
		\right)
\eeq
Here $Q_u=2/3(-1/3)$ for q=u(d) and $R_q=v_1/v_2^* (v_2/v_1^*)$ 
for q=u(d), 
and $m_q$ is the quark mass.
The squark matrix is diagonalized by  $D_{qij}$ such that 
\beq
D_{q}^\dagger M_{\tilde{q}}^2 D_q={\rm diag}(M_{\tilde{q}1}^2,
              M_{\tilde{q}2}^2)
\eeq
As mentioned in the introduction the relative velocities of the LSP 
hitting the targets are small, and consequently we can approximate 
the effective
interaction governing the neutralino-quark scattering by an   
effective four-fermi interaction. We give now the result of our 
analysis including all the relevant CP violating effects  in a  
softly broken MSSM. The effective four fermi 
 interaction is given by
\beqn
{\cal L}_{eff}=\bar{\chi}\gamma_{\mu} \gamma_5 \chi \bar{q}
\gamma^{\mu} (A P_L +B P_R) q+ C\bar{\chi}\chi  m_q \bar{q} q
+D  \bar{\chi}\gamma_5\chi  m_q \bar{q}\gamma_5 q\nonumber\\
+E\bar{\chi}i\gamma_5\chi  m_q \bar{q} q
+F\bar{\chi}\chi  m_q \bar{q}i\gamma_5 q
\eeqn
where our choice of the metric is $\eta_{\mu\nu}=(1,-1,-1,-1)$. 
The deduction of Eq.(6) starting from the microscopic supergravity 
Lagrangian is given in Appendix A. Here we give the results. 
The first two terms ($A$, $B$) are spin-dependent interactions and arise
from the $Z$ boson  and the sfermion exchanges. For these our analysis gives
\beq
A=\frac{g^2}{4 M^2_W}[|X_{30}|^2-|X_{40}|^2][T_{3q}-
e_q sin^2\theta_W]
-\frac{|C_{qR}|^2}{4(M^2_{\tilde{q1}}-M^2_{\chi})}
-\frac{|C^{'}_{qR}|^2}{4(M^2_{\tilde{q2}}-M^2_{\chi})}
\eeq
\beq
B=-\frac{g^2}{4 M^2_W}[|X_{30}|^2-|X_{40}|^2]
e_q sin^2\theta_W +
\frac{|C_{qL}|^2}{4(M^2_{\tilde{q1}}-M^2_{\chi})}
+\frac{|C^{'}_{qL}|^2}{4(M^2_{\tilde{q2}}-M^2_{\chi})}
\eeq
where 
\beq
C_{qL}=\sqrt{2} (\alpha_{q0} D_{q11} -\gamma_{q0} D_{q21})
\eeq
\beq
C_{qR}=\sqrt{2} (\beta_{q0} D_{q11} -\delta_{q0} D_{q21})
\eeq
\beq
C^{'}_{qL}=\sqrt{2} (\alpha_{q0} D_{q12} -\gamma_{q0} D_{q22})
\eeq
\beq
C^{'}_{qR}=\sqrt{2} (\beta_{q0} D_{q12} -\delta_{q0} D_{q22})
\eeq
and where $\alpha$, $\beta$, $\gamma$, and $\delta$ are given by\cite{falk4}

\beq
\alpha_{u(d)j}=\frac{g m_{u(d)}X_{4(3)j}}{2 m_W\sin\beta(\cos\beta)}
\eeq

\beq
\beta_{u(d)j}=eQ_{u(d)j}X'^{*}_{1j}+\frac{g}{cos\theta_W}X'^{*}_{2j}
(T_{3u(d)}-Q_{u(d)}\sin^2\theta_W)
\eeq

\beq
\gamma_{u(d)j}=eQ_{u(d)j}X'_{1j}-
\frac{gQ_{u(d)}\sin^2\theta_W}{cos\theta_W}X'_{2j}
\eeq

\beq
\delta_{u(d)j}= \frac{- g m_{u(d)}X^*_{4(3)j}}{2 m_W\sin\beta(\cos\beta)}
\eeq
Here g is the $SU(2)_L$ gauge coupling and 
\beq
X'_{1j}=X_{1j}\cos\theta_W+X_{2j}\sin\theta_W
\eeq
\beq
X'_{2j}=-X_{1j}\sin\theta_W+X_{2j}\cos\theta_W
\eeq
The effect of the CP violating phases enter via the neutralino
eigen-vector components $X_{ij}$
 and via the matrix $D_{qij}$
that diagonalizes the squark mass matrix.

 The C term in Eq.(6) represents the scalar interaction which gives rise to 
 coherent scattering.. It receives contributions from the sfermion 
 exchange, from the CP even light Higgs ($h^0$) exchange, and from the
 CP even heavy Higgs ($H^0$) exchange. Thus   
 \begin{equation}
  C=C_{\tilde{f}}+C_{h^0}+C_{H^0}
  \end{equation}
   where 
\beq
C_{\tilde{f}}(u,d)= -\frac{1}{4m_q}\frac{1}
{M^2_{\tilde{q1}}-M^2_{\chi}} Re[C_{qL}C^{*}_{qR}]
-\frac{1}{4m_q}\frac{1}
{M^2_{\tilde{q2}}-M^2_{\chi}} Re[C^{'}_{qL}C^{'*}_{qR}]
\eeq

\beqn
C_{h^0}(u,d)=-(+)\frac{g^2}{4 M_W M^2_{h^0}}
\frac{\cos\alpha(sin\alpha)}{\sin\beta(cos\beta)} Re\sigma
\eeqn

\beqn
C_{H^0}(u,d)=
\frac{g^2}{4 M_W M^2_{H^0}}
\frac{\sin\alpha(cos\alpha)}{\sin\beta(cos\beta)} Re \rho
\eeqn
Here $\alpha$ is the Higgs mixing angle,  
(u,d) indicate the flavor of the quark
involved in the scattering, and  $\sigma$ and $\rho$ are 
defined by

\beq
\sigma= 
 X_{40}^*(X_{20}^*-\tan\theta_W X_{10}^*)\cos\alpha
+X_{30}^*(X_{20}^*-\tan\theta_W X_{10}^*)\sin\alpha
\eeq

\beq
\rho=
- X_{40}^*(X_{20}^*-\tan\theta_W X_{10}^*)\sin\alpha
+X_{30}^*(X_{20}^*-\tan\theta_W X_{10}^*)\cos\alpha
\eeq
The last three terms D,E and F in eq.(6) are given by 
\beq
D(u,d)= C_{\tilde{f}}(u,d)
+\frac{g^2}{4M_W}\frac{cot\beta(tan\beta)}{m_{A_0}^2}Re\omega
\eeq
\beqn
E(u,d)=T_{\tilde{f}}(u,d)+
\frac{g^2}{4M_W} [
-(+)  \frac{cos\alpha(sin\alpha)}
{\sin\beta(cos\beta)}  \frac{Im\sigma} {m_{h^0}^2}   
+\frac{sin\alpha(cos\alpha)}
{\sin\beta(cos\beta)}\frac{Im\rho}{m_{H^0}^2}  ]
\eeqn

\beqn
F(u,d)=T_{\tilde{f}}(u,d)
+\frac{g^2}{4M_W} \frac{cot\beta(tan\beta)}{m_{A^0}^2}
Im\omega
\eeqn
where $A^0$ is the CP odd Higgs and  where $\omega$ is defined by 
\beq 
 \omega= 
 -X_{40}^*(X_{20}^*-\tan\theta_W X_{10}^*)\cos\beta
+X_{30}^*(X_{20}^*-\tan\theta_W X_{10}^*)\sin\beta
\eeq
and 
\beq
T_{\tilde{f}}(q)= \frac{1}{4m_q}\frac{1}
{M^2_{\tilde{q1}}-M^2_{\chi}} Im[C_{qL}C^{*}_{qR}]
+\frac{1}{4m_q}\frac{1}
{M^2_{\tilde{q2}}-M^2_{\chi}} Im[C^{'}_{qL}C^{'*}_{qR}]
\eeq
In the limit of vanishing CP violating phases our 
results of A, B and C
limit to the result of reference\cite{jungman}. 
In that limit we have a difference
of a minus sign in the Z-exchange terms of Ref. \cite{drees} in
 their equations (2a, 2b, A1). Further, in the same limit of 
 vanishing CP violating phases our results go to those of Ref. \cite{flores}
with an overall minus sign difference in the three exchange terms, i.e.,
the Z, the sfermions and the higgs terms.
(see Appendix B for details).
To compare our results to those of 
reference\cite{falk3} we note that the analysis of reference\cite{falk3}
 was limited to 
the case of two CP violating phases and it gave the analytic results 
for only some of the co-efficients in the low energy expansion of 
Eq.(6). 
  The terms E and F given by
Eqs.(26) and (27) are new and vanish in the limit when CP is conserved.
The term D given by Eq.(25) is non-vanishing in the limit when CP 
phases vanish. However, this term is mostly ignored in the literature
as  its contribution is suppressed because of the small velocity
of the relic neutralinos. In fact the contributions of D,E and F are
expected to be relatively small and we ignore them in our numerical 
analysis here.

For the computation of the event rates from nuclear 
targets in the direct detection of dark matter we follow the 
techniques discussed in Ref.\cite{arno} 
and we refer the reader to it for details. 
We give now the numerical estimates of the  CP violating
effects on event rates. First we consider the case when
the EDM constraint is not imposed. In Fig.1 we exhibit the ratio
$R/R(0)$ for Ge where R is the event rate with CP violation arising from
a non-vanishing $\theta_{\mu}$ and R(0) is the 
event rate in the absence of CP violation. The figure illustrates that
the effect of the CP violating phase can be very large. In fact, as can be
seen from Fig.1 the variations can be as large as 2-3 orders of 
magnitude. However, as noted above the EDM constraint was not imposed
here.  
Next we give the analysis with inclusion of  
the EDM constraints. 
 For this purpose we work in the parameter space 
where the cosmological relic density constraint and the EDM
constraints are simultaneously satisfied and  we compute the 
ratio R/R(0) for Ge in this region. Specifically the satisfaction of the relic 
density and the EDM constraints is achieved by varying the
parameters of the theory.   The satisfaction of the EDM
constraint is achieved through the cancellation mechanism discussed
in Refs.\cite{in1,in2}. The result 
of this analysis is exhibited in Fig.2. Here we find the range of
variation of R/R(0) with $\theta_{\mu}$ to be much smaller although 
still substantial. Thus from Fig.2 we find that
the variation of R/R(0) has a range of up to a factor of 2 over 
most of the allowed parameter space satisfying the relic density and
the EDM constraints in the regions of the parameter investigated.
This variation is substantially smaller than the one observed in Fig.1 
when the EDM constraints were not imposed.

In conclusion, we have given in this paper the first complete 
analytic analysis of the effects of CP violating phases on the
event rates in the direct detection of dark matter within the
framework of MSSM with no generational mixing. We find that the
CP violating effects can generate variations in the event rates
up to 2-3 orders of magnitude. However, with the inclusion of the
EDM constraints the effects are much reduced although still
significant in that  
the variations could be up to a factor $\sim 2$
as seen from the analysis over the region of the  parameter 
space investigated. Of course the parameter space of MSSM
is quite large and there may exist other regions of the parameter
space of MSSM where the CP violating effects on event rates 
consistent with the EDM constraints are even larger. 

\noindent 
 This research was supported in part by NSF grant 
PHY-96020274.

\begin{figure}
\begin{center}
\includegraphics[angle=0,width=2.5in]{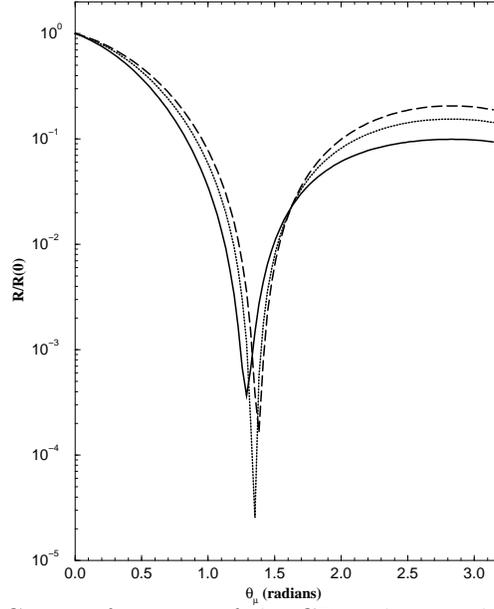}
\caption{Plot of R/R(0) for Ge as a function of the CP violating phase
$\theta_{\mu}$ in the MSSM case when tan$\beta$=2, $m_0$=100 GeV,
$|A_0|$=1 for the cases when the gluino mass is 500 GeV
(solid curve), 600 GeV (dotted curve), and 700 GeV (dashed
curve)}
\label{fig1}
\end{center}
\end{figure}

\begin{figure}
\begin{center}
\includegraphics[angle=0,width=2.5in]{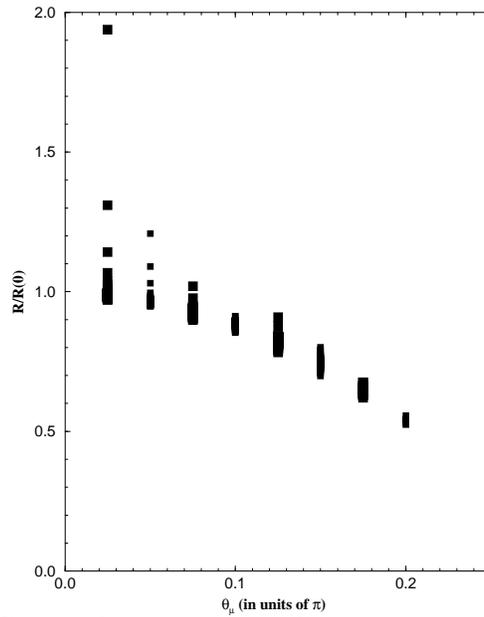}
\caption{Plot of R/R(0) for Ge as a function of the CP violating phase
$\theta_{\mu}$ for the parameter space discussed in the
text. The parameter space spans regions satisfying
the relic density and the EDM constraints obtained by varying other
parameters in the theory.
}
\label{fig2}
\end{center}
\end{figure}

\newpage
\section{Appendix A}
In this appendix we give a derivation  of  the four fermi
neutralino-quark effective 
Lagrangian with CP violating phases given in the text.

\subsection{squark exchange terms}
From the microscopic Lagrangian of quark-squark-neutralino\cite{in1}
\beq
{-\cal L}=\bar{q}[C_{qL}P_L+C_{qR}P_R]\chi\tilde{q_1}+
\bar{q}[C^{'}_{qL}P_L+C^{'}_{qR}P_R]\chi\tilde{q_2}+H.c.
\eeq
the effective lagrangian for $q-\chi$ scattering via the
exchange of squarks is given by 
\beqn
{\cal L}_{eff}=\frac{1}{M^{2}_{\tilde{q_1}}-M^{2}_{\chi}}
\bar{\chi}[C^{*}_{qL}P_R+C^{*}_{qR}P_L]q\bar{q}[C_{qL}P_L+C_{qR}P_R]
\chi\nonumber\\
+\frac{1}{M^{2}_{\tilde{q_2}}-M^{2}_{\chi}}
\bar{\chi}[C^{*'}_{qL}P_R+C^{*'}_{qR}P_L]q
\bar{q}[C^{'}_{qL}P_L+C^{'}_{qR}P_R]
\chi
\eeqn
We use Fierz reordering to write the Lagrangian in terms
of the combinations 
$\bar{\chi}\chi\bar{q}q$, $\bar{\chi}\gamma_5\chi\bar{q}\gamma_5q$,
$\bar{\chi}\gamma^{\mu}\gamma_5\chi\bar{q}\gamma_{\mu}q$,$
\bar{\chi}\gamma^{\mu}\gamma_5\chi\bar{q}\gamma_{\mu}\gamma_5q$,
$\bar{\chi}\gamma_5\chi\bar{q}q$ and $\bar{\chi}\chi\bar{q}\gamma_5q$.
For this purpose, we define the 16 matrices
\beq
\Gamma^A=\{1,\gamma^0,i\gamma^i,i\gamma^0\gamma_5,
\gamma^i\gamma_5,\gamma_5,i\sigma^{0i},\sigma^{ij}\}: ~~i,j=1-3
\eeq
with the normalization
\beq
tr(\Gamma^A\Gamma^B)=4\delta^{AB}
\eeq
The Fierz rearrangement formula with the above definitions and 
normalizations is
\beq
(\bar{u_1}\Gamma^Au_2)(\bar{u_3}\Gamma^Bu_4)=\sum_{C,D}F^{AB}_{CD}
(\bar{u_1}\Gamma^Cu_4)(\bar{u_3}\Gamma^Du_2)
\eeq
where $u_j$ are Dirac or Majorana spinors and
\beq
F^{AB}_{CD}=
-(+)\frac{1}{16}tr(\Gamma^C\Gamma^A\Gamma^D\Gamma^B)
\eeq
and where the +ve sign is for commuting u spinors and the -ve sign is for
the anticommuting u fields. In our case we have to use the -ve sign since
we are dealing with quantum Majorana and Dirac fields in the Lagrangian.
 We give below the Fierz rearrangement of the four combinations that appear
  in Eq.(31) above:
\beqn
\bar{\chi}q\bar{q}\chi=-\frac{1}{4}\bar{\chi}\chi\bar{q}q-\frac{1}{4}
\bar{\chi}\gamma_5\chi\bar{q}\gamma_5q+\frac{1}{4}\bar{\chi}\gamma^{\mu}
\gamma_5\chi\bar{q}\gamma_{\mu}\gamma_5q\nonumber\\
\bar{\chi}\gamma_5
q\bar{q}\chi=\frac{1}{4}\bar{\chi}\gamma^{\mu}\gamma_5
\chi\bar{q}\gamma_{\mu}
q-\frac{1}{4}
\bar{\chi}\chi\bar{q}\gamma_5q-\frac{1}{4}\bar{\chi}\gamma_5
\chi\bar{q}q\nonumber\\
\bar{\chi}
q\bar{q}\gamma_5
\chi=-\frac{1}{4}\bar{\chi}\gamma^{\mu}\gamma_5
\chi\bar{q}\gamma_{\mu}
q-\frac{1}{4}
\bar{\chi}\chi\bar{q}\gamma_5q-\frac{1}{4}\bar{\chi}\gamma_5
\chi\bar{q}q\nonumber\\
\bar{\chi}\gamma_5
q\bar{q}\gamma_5
\chi=-\frac{1}{4}\bar{\chi}\chi\bar{q}q-\frac{1}{4}
\bar{\chi}\gamma_5\chi\bar{q}\gamma_5q-\frac{1}{4}\bar{\chi}\gamma^{\mu}
\gamma_5\chi\bar{q}\gamma_{\mu}\gamma_5q
\eeqn
In the above we have used the metric $\eta_{\mu\nu}=(1,-1,-1,-1)$, 
 and we also used the fact that $\chi$ is a 
Majorana
field so that  $\bar{\chi}\gamma_{\mu}\chi=0$
and $\bar{\chi}\sigma_{\mu\nu}\chi=0$.
By rearranging the terms to be in the form of ${\cal L}_{eff}$
of Eq.(6) we can find out
directly
the squark contributions to $A, B, C, D, E$ and $F$ terms as given in
the text of the  paper.

\subsection{Z boson exchange}

From the $q-Z-q$ and $\chi-Z-\chi$ interactions
in Eqs. (c62) and (c87a) of Ref.\cite{kane}, we obtain the
following effective Lagrangian for the $q-\chi$ scattering via Z-exchange:
\beq
{\cal L}_{eff}=\frac{g^2}{4M^2_z \cos^2\theta_W}
[|X_{30}|^2-|X_{40}|^2]\bar{q}\gamma^{\mu}[d_{qL}P_L+
d_{qR}P_R]q\bar{\chi}\gamma_{\mu}\gamma_5\chi
\eeq
where $d_{qL}=T_{3L}-e_q\sin^2\theta_W$ and $d_{qR}=-e_q\sin^2\theta_W$.
From Eq.(37) we can read off the
 contribution to A and B from the Z exchange. These contributions are
 given in Eqs.(7) and (8). 

\subsection{Higgs exchange terms}
Higgs exchange will contribute to $C, D, E$ and $F$ terms. From the
interaction Lagrangian of ${\cal L}_{H\chi\chi}$ and
${\cal L}_{Hq\bar{q}}$ 
in Eqs. (4.47) and (4.10) respectively of Ref.\cite{haber},
one can get the effective Lagrangian for $q-\chi$ scattering via $h^0$,
$H^0$ and $A^0$ exchanges. In our formalism we use $h^0$, $H^0$ and
$A^0$ for the light, heavy and CP-odd neutral higgs.
There are six contributions: three higgs exchange terms
for the up flavor and three for the down flavor. To illustrate we
 choose the up quark scattering with $\chi$ via  the
exchange  of the heavy higgs $H^0$ ($H^0_1$ in the notation of
Ref.\cite{haber}):
\beq
{\cal L}_{eff}=\frac{1}{m^2_{H0}}(J^1_{H0}+J^2_{H0})I^u_{H0}
\eeq
where
\beqn
J^1_{H0}=-\frac{g}{2}\cos\alpha\bar{\chi}
(Q^{"*}_{00}P_L+Q^{"}_{00}P_R)\chi\nonumber\\
J^2_{H0}=\frac{g}{2}\sin\alpha\bar{\chi}
(S^{"*}_{00}P_L+S^{"}_{00}P_R)\chi\nonumber\\
I^u_{H0}=-\frac{gm_u\sin\alpha}{2M_w \sin\beta}\bar{u}u
\eeqn
where $Q_{00}^{''*}, S_{00}^{''*}$ are as defined in Ref.\cite{haber}.
Defining $\rho$ by 
\beq
\rho=Q^{"}_{00}\cos\alpha-S^{"}_{00}\sin\alpha
\eeq
we get the $H^0$ contribution to ${\cal L}_{eff}$:
\beqn
{\cal L}_{eff}=\frac{g^2m_u\sin\alpha}{4M_W m^2_{H0}\sin\beta}Re(\rho)
\bar{\chi}\chi\bar{u}u\nonumber\\
+\frac{ig^2m_u\sin\alpha}{4M_Wm^2_{H0}\sin\beta}Im(\rho)
\bar{\chi}\gamma_5\chi\bar{u}u
\eeqn
From Eq.(41) we can read off directly the contributions 
$C_{H0}(u) $ and $E_{H0}(u)$ as given by Eqs.(22) and (26).

\newpage
\section{Appendix B}
Here we compare our results with those of Ref.\cite{flores}
which is in the limit of no CP violation, no sfermion mixing
and no heavy Higgs. In the limit of no CP violation and no sfermion
mixing 
$C_L, C_L', C_R, C_R'$ given by Eqs.(9-12) in the text reduce to the 
following:

\begin{eqnarray}
C_L=\sqrt{2} \alpha_{q0},~~C_L'=-\sqrt 2 \gamma_{q0}\nonumber\\
C_R=\sqrt {2} \beta_{q0}, ~~C_R'=-\sqrt{2} \delta_{q0}
\end{eqnarray}
To express the above in the notation of Ref.\cite{flores}
we set $X_{10}^*=\beta$, $X_{20}^*=\alpha$, $X_{30}^*=\delta$,
and $X_{40}^*=\gamma$ keeping in mind that the $H_1$ and $H_2$ of
Ref.\cite{flores} are defined oppositely to our notation. 
Using the above along with Eqs.(13-16) we find for 
$T_3=\frac{1}{2}(-\frac{1}{2})$ 

\begin{eqnarray}
|C_L|^2=\frac{m_u^2(m_d^2)\gamma^2(\delta^2)}{\nu_1^2(\nu_2^2)}\nonumber\\
|C_L'|^2=2(g_1\frac{1}{2}Y_R\beta)^2\nonumber\\
|C_R|^2=2(\alpha g_2 T_{3}+\beta g_1\frac{Y_L}{2})^2\nonumber\\
|C_R'|^2=\frac{m_u^2(m_d^2)\gamma^2(\delta^2)}{\nu_1^2(\nu_2^2)}
\end{eqnarray}
where $Y_L$ is the hypercharge,
$\nu_1=<H_1>$, and $\nu_2=<H_2>$ and where $H_1$ and $H_2$ are in
the notation of Ref.\cite{flores}. 
Further using the identity

\begin{equation}
\frac{g_2^2}{cos^2\theta_W}(T_{3}-e_qsin^2\theta_W)=
-(g_1 sin\theta_W+g_2 cos\theta_W)(\frac{1}{2}Y_Lg_1 sin\theta_W-
T_3g_2 cos\theta_W)
\end{equation}
where the left hand side of Eq.(44) is written in the
form used in  Ref.\cite{flores}, 
 we can express A and B in the limit of no CP violation and
 no sfermion  mixing as follows: for $T_3=\frac{1}{2}(-\frac{1}{2})$

\begin{eqnarray}
A=\frac{(\gamma^2-\delta^2)}{4M_Z^2}
(g_1 sin\theta_W+g_2 cos\theta_W)(\frac{1}{2}Y_L g_1 sin\theta_W-
T_3g_2 cos\theta_W)\nonumber\\
-\frac{(\alpha g_2 T_{3}+\beta g_1\frac{Y_L}{2})^2}
{2(M_{\tilde q_L}^2-M_{\chi}^2)}
-\frac{m_u^2(m_d^2)\gamma^2(\delta^2)}
{4(M_{\tilde q_R}^2-M_{\chi}^2)\nu_1^2(\nu_2^2)} 
\end{eqnarray}
\begin{eqnarray}
B=\frac{(\gamma^2-\delta^2)}{4M_Z^2}
(g_1 sin\theta_W+g_2 cos\theta_W)(\frac{1}{2}Y_L g_1 sin\theta_W-
T_3g_2 cos\theta_W)\nonumber\\
+\frac{(g_1\frac{1}{2}Y_R\beta)^2}{2(M_{\tilde q_R}^2-M_{\chi}^2)}
+\frac{m_u^2(m_d^2)\gamma^2(\delta^2)}
{4(M_{\tilde q_R}^2-M_{\chi}^2)\nu_1^2(\nu_2^2)} 
\end{eqnarray}
To compare  our $C$ term with that of Ref.\cite{flores} we
again go to the limit of vanishing CP violating phases, 
assume no sfermion mixing, and in addition ignore
the heavy Higgs exchange contribution (i.e., the term $C_{H^0}$
of Eq.(22) in the text).
Then using similar notational changes as above we find that
our $C$ under the approximations made in Ref.\cite{flores}
is given by
\begin{eqnarray}
C=\frac{g_2^2}{4M_W m_{h^0}^2}\frac{(-cos\alpha)(sin\alpha)}
{(sin\beta)(cos\beta)}(\alpha-\beta tan\theta_W)(\gamma cos\alpha
+\delta sin\alpha)\nonumber\\
-\frac{g_2}{4M_W}(\frac{\alpha g_2 T_{3}+\beta g_1\frac{Y_L}{2}}
{M_{\tilde q_L}^2-M_{\chi}^2}-\frac{\beta g_1\frac{Y_R}{2}}
{M_{\tilde q_R}^2-M_{\chi}^2}) 
\frac{\gamma (\delta)}{sin\beta(cos\beta)} ;~~T_3=\frac{1}{2}(-\frac{1}{2})
\end{eqnarray}
Comparing our results for A,B and C with those of Ref.\cite{flores}
we find that our Z, sfermion and Higgs exchange terms have an
overall minus sign relative to those of Ref.\cite{flores}.


\begin{thebibliography}{99}
\bibitem{correction}
The following corrections should be included in Refs.\cite{in1,in2}:
In the first paper of Ref.\cite{in1} a minus sign should be inserted
on the right hand side of Eq.(16). In the expressions of $\eta_{qik}$
(or $\eta_{fik}$) in Refs.\cite{in1,in2} 
the first $\kappa_q$ (or $\kappa_f$) should be preceded by a minus sign.
\end{thebibliography}
\end{document}